\documentstyle[aps,prl,epsfig]{revtex}

\def\var{{\rm var}(\Delta\!E)}
\def\DE{\Delta\!E}
\def\pG{p(\Gamma)}
\def\Pqm{P_{\rm qm}(t)}
\def\tqm{t^{*}}
\def\Gqm{\Gamma^{*}}
\def\ns{n^{*}}
\def\nss{n^{\dagger}}
\def\t0{t_{0}}
\def\tH{\tau_{\!\raisebox{-0.6ex}{{\footnotesize H}}}}
\def\heff{\hbar_{\rm ef\/f}}
\begin{document}
\draft
\twocolumn[\hsize\textwidth\columnwidth\hsize\csname @twocolumnfalse\endcsname

\title{Conductance Fluctuations of Generic Billiards: Fractal or Isolated?}

\author{L.~Hufnagel, R.~Ketzmerick, and M.~Weiss}
\address{ Max-Planck-Institut f\"ur Str\"omungsforschung
and Institut f\"ur Nichtlineare Dynamik der Universit\"at G\"ottingen,\\
Bunsenstr.~10, 37073 G\"ottingen, Germany}

\date{\today}

\maketitle

\begin{abstract}
We study the signatures of a classical mixed phase space for open quantum systems.
We find the scaling of the break time up to which quantum mechanics
mimics the classical staying probability and derive the distribution of resonance widths.
Based on these results we explain why for mixed systems two types of conductance fluctuations
were found:
quantum mechanics divides the hierarchically structured chaotic component of
phase space into two parts \-- one yields fractal conductance fluctuations while
the other causes isolated resonances.
In general, both types appear together, but on different energy scales.
\end{abstract}


]

%------------------------------------------------------

\begin{figure}
\begin{center}
    \epsfxsize=7.6cm
    \leavevmode
    \epsffile{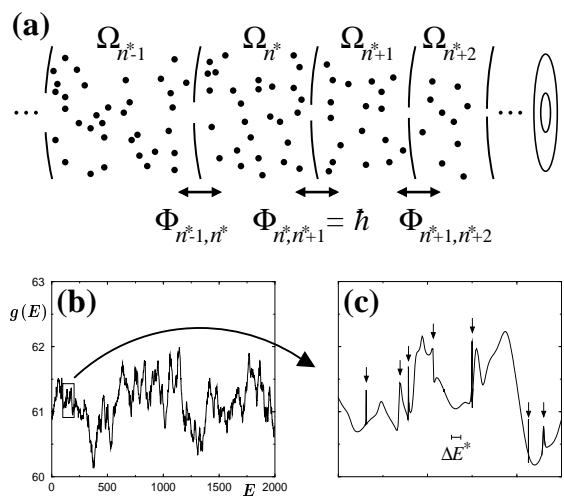}
\caption[model]{
(a) Sketch of a Poincar\'e surface of section,
showing the hierarchical structure of the chaotic component of phase space
in the vicinity of a regular island.
The quantum flux barrier at $\ns$ divides this hierarchy into two parts.
(b) Dimensionless conductance vs. energy (in units of the mean level spacing) for a quantum graph $(v = 32)$
showing fractal fluctuations on large energy scales and (c) isolated resonances (arrows)
on scales  smaller than $\DE^*$. 
}
\label{modelfig}
\end{center}
\end{figure}

Generic Hamiltonian systems are nonintegrable and have a mixed phase space, where regions
of regular and chaotic motion coexist\cite{MM74,lich92}.
The chaotic dynamics in mixed systems is clearly distinct from the fully chaotic case.
For an open system this is manifested  in a power-law decay of the staying probability\cite{powerlaw}
\begin{equation}
\label{clpot}
P(t)\sim t^{-\gamma},\quad \gamma>1\quad,
\end{equation}
in contrast to the typically exponential decay in fully chaotic systems.
The power law originates from partial transport barriers\cite{kay84}, e.g., Cantori, dividing
the chaotic part of phase space into an infinite hierarchy of partially connected chaotic
regions (for a sketch see Fig.~\ref{modelfig}a).
These regions of ever decreasing size are connected by fluxes, i.e., exchanged phase space volumes 
on the energy surface per time.
It is a central question of quantum chaos, how the hierarchical structure and the dynamics of 
a generic classical phase space show up in quantum properties.
An important fact is that quantum dynamics drastically differs from classical dynamics
once the classical flux between connected regions becomes smaller than the Planck constant\cite{bohigas}.
For the hierarchy of chaotic regions connected by decreasing fluxes 
this introduces a quantum flux barrier.
It divides the chaotic part of phase space into two parts with different quantum properties.
For {\it closed} quantum systems this leads to chaotic states before and
hierarchical states behind the flux barrier\cite{ketz00}.
For {\it open} quantum systems, the search for signatures of the hierarchical phase space
has concentrated on conductance fluctuations, a central phenomenon of mesoscopic physics\cite{datta95,jala00}.
They occur as a function of an external parameter, e.g., magnetic field or energy,
when the phase coherence length exceeds the sample size.
A semiclassical analysis led from Eq.~(\ref{clpot}) to the prediction of fractal conductance
fluctuations (FCF)\cite{ketz96}.
For the typical $\gamma<2$, they are characterized by a fractal dimension $D=2-\gamma/2$
of the conductance curve $g(E)$ and by a scaling of the variance of conductance increments
\begin{equation}
\label{variance}
\var \equiv \langle (g(E+\DE) - g(E))^2 \rangle_{E} \sim |\DE|^\gamma\quad,
\end{equation}
on energy scales $\DE$, which must be larger than the mean level spacing.
In fact, FCF have been found in experiments on gold wires\cite{hegg96},
semiconductor nanostructures\cite{sach98} and numerically\cite{ital00,taka00}.
Surprisingly, a new type of conductance fluctuations has
been found numerically for the cosine billiard with a mixed phase space\cite{bodo00}.
In contrast to the expected FCF, the conductance $g(E)$ shows a
smoothly varying background with many isolated resonances.
These narrow resonances do not lead to fractal properties. 
They are reflected, however, in the variance of conductance increments
where a power law $\var\sim(\DE)^\delta$ {\it below} the mean level spacing was observed.
The exponent $\delta$ appearing in this quantum regime seemed to coincide with the
classical exponent $\gamma$ from Eq.~(\ref{clpot}), contradicting semiclassical intuition.
Neither the origin of these isolated resonances nor the observed power law have found an
explanation so far.
Even the fundamental riddle is unresolved:
Why are there two types of conductance fluctuations in mixed systems?
In this paper, we derive and numerically verify the
scaling of the break time $t^*$ until which the quantum 
staying probability mimics the classical power-law decay of Eq.~(\ref{clpot}).
We derive the resonance width distribution $p(\Gamma)$, the main characteristic of an open quantum system,
for the states behind the flux barrier.
We then apply these results to conductance fluctuations and show
that the phase space regions before the flux barrier give rise to FCF, while the regions behind
are the origin of isolated resonances. 
Thus FCF and isolated resonances will in general appear together (Fig.~\ref{modelfig}b,c).
The smallest energy scale of FCF is shown to be $\DE^*=h/t^*$, below which isolated resonances appear.
They are characterized by an asymptotic power law for the variance of conductance
increments $\var\sim\DE$ for $\DE \rightarrow 0$.
Furthermore, these results allow to predict which type of conductance fluctuations will
dominate in a given numerical or experimental setup.
We thus unify the contradicting findings of FCF
\cite{ketz96,hegg96,sach98,ital00,taka00} and isolated resonances\cite{bodo00}.
%
% ------------------- end intro ------------------------

% classical chain model
%
We will use the simplest model describing the infinite hierarchy of partially connected chaotic
regions\cite{HCM85}.
It is a chain of regions $n=0,1,\dots$ with downscaling
volumes $\Omega_n=\Omega_0\omega^n$ ($\omega<1$) on the energy surface.
Neighboring regions are connected by decreasing fluxes $\Phi_{n,n+1}=\Phi_{0,1}\varphi^n$
($\varphi<\omega$), as sketched in Fig.~\ref{modelfig}a.
In the presence of a flux $\Phi>\Phi_{0,1}$ for leaving the chain from region $n=0$ the
staying probability in the chain, when started in region $n=0$, decays according to
Eq.~(\ref{clpot}) with $\gamma=1/(1-\ln\omega/\ln\varphi)$\cite{HCM85}.
%

% quantum restictions of the cl. chain model
%
Quantum dynamics in a $d$-dimensional system can mimic the classical flux between two regions
if it is larger than $\hbar^{d-1}$ while in the opposite case the regions are
coupled perturbatively\cite{bohigas}.
Applying this idea to the chain model yields the position $\ns$ of the flux barrier
\begin{equation}
\Phi_{\ns,\ns+1}\approx\hbar^{d-1}\quad,
\end{equation}
which divids the chain into two parts:
the part before the flux barrier $(n<\ns)$ where quantum mechanics mimics classical dynamics and
the regions with $n>\ns$ which are coupled perturbatively to each other.
The position of the flux barrier is $\ns=\ln\heff/\ln\varphi$ with $\heff\equiv\hbar^{d-1}/\Phi_{0,1}$ 
being the effective Planck constant of the chain model\cite{volume}.
%

\begin{figure}
\begin{center}
\epsfig{figure=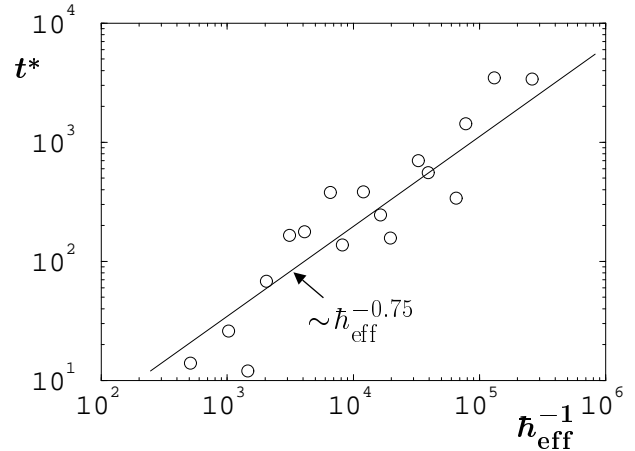,width=8.5cm}\hfill
\caption[]{The quantum break time $\tqm$ up to which $\Pqm\approx P(t)$ (with deviation $\le 20\%$)
vs. $\heff^{-1}$ for the separatrix map ($\gamma = 1.33$, parameters as in Ref.~\cite{ital00}).
The solid line corresponds to the prediction of Eq.~(\ref{tqm}) using $\tau_H\sim\heff^{-1}$ for the
separatrix map.
}
\label{tqmfig}
\end{center}
\end{figure}

%
% break time
The flux barrier introduces an important new time scale $\tqm=\Omega_{\ns+1}/\Phi_{\ns,\ns+1}$.
Beyond this time, regions with $n>\ns$ are explored and quantum dynamics has to differ from
classical dynamics.
Therefore, at most up to this break time the quantum mechanical staying probability
$\Pqm$ can follow the classical power-law decay.
The time $\tqm$ scales as
\begin{equation}\label{tqm}
\tqm\sim \tH\cdot\heff^{1-1/\gamma}\quad,
\end{equation}
where $\tH=h/\Delta$ is the Heisenberg time and $\Delta=h^{d}/\sum_n\Omega_{n}$ is the
mean level spacing.
This result is in contrast to previous predictions for the break time\cite{casa99,sund99}, 
but agrees with the numerical findings for the kicked rotor at kicking strength $K=2.5$,
where $\gamma\approx 2$\cite{casa99}.
The scaling of the break time according to Eq.~(\ref{tqm}) is confirmed over three
orders of magnitude in $\heff$ for the separatrix map\cite{ital00} with $\gamma=1.33$ (Fig.~\ref{tqmfig}).
%

\begin{figure}
\begin{center}
\epsfig{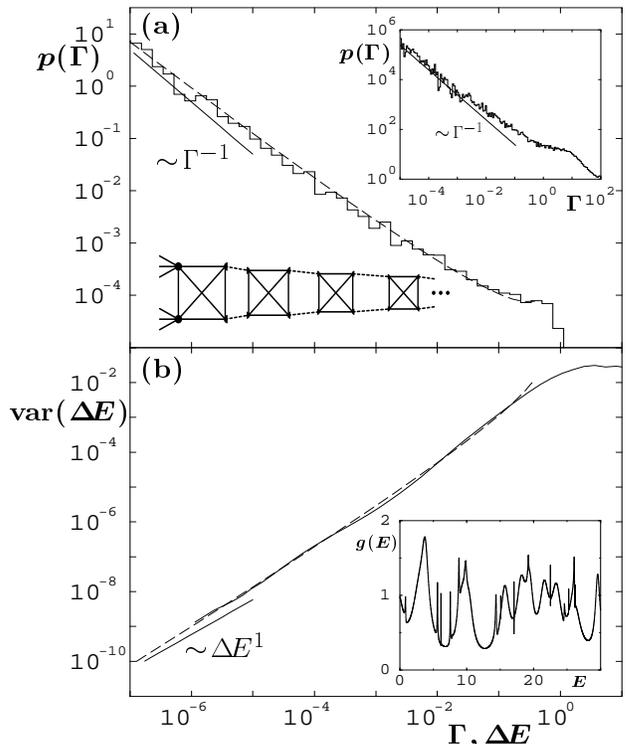}\hfill
\caption{
(a) Distribution of resonance widths (solid line) for a quantum graph with $v=4$ (lower inset)
and the prediction of Eq.~(\ref{pertgamma}) (dashed line).
The slow convergence to the asymptotic $1/\Gamma$ can be seen.
The distribution for the kicked rotor ($K=2.5$, $\hbar=2\pi/15000$)
shows the same behavior (upper inset).
(b) The variance of conductance increments shows the asymptotic $\var\sim\DE$
and a broad transition region in agreement with the expected behavior
derived from Eq.~(\ref{pertgamma}) ($\var\approx p(\Gamma=\DE)\cdot(\DE)^2$, dashed line).
The inset shows the corresponding conductance with many isolated resonances.
Energies are in units of the mean level spacing.
}
\label{pogfig}
\end{center}
\end{figure}

% perturbation theory
%
A major characteristic of open quantum systems is the distribution $\pG$ of
resonance widths.
Since the states behind the flux barrier couple only weakly to the part before the flux barrier 
their distribution of resonance widths can be calculated perturbatively.
The typical resonance width $\Gamma_n$ of the states localized in region $n>\ns+1$ is
proportional to the product of all the individual couplings connecting the regions from $\ns+1$ to $n$.
This yields
\begin{equation}\label{typgamma}
\frac{\Gamma_n}{\Gqm} =\!\!\!\prod_{j=\ns+1}^{n-1}\frac{\Phi_{j,j+1}}{\hbar^{d-1}}
=\!\!\!\prod_{j=\ns+1}^{n-1}\varphi^{j-\ns} \approx \varphi^{(n-\ns)^2/2}\quad,
\end{equation}
where $\Gqm=h/\tqm$ and we assumed $n-\ns\gg 1$ for the final step.
Using the cumulative distribution of resonance widths
$P_{int}(\Gamma_n/\Gqm)= \int_{0}^{\Gamma_n/\Gqm}\;\pG d\Gamma\sim\sum_{j=n}^{\infty}\Omega_j\sim\omega^n$
one finds
\begin{equation}\label{pertgamma}
\pG\sim\frac{1}{\Gamma}
\cdot\frac{ \exp\left(\ln\omega\sqrt{ 2\ln{(\Gamma/\Gqm)}/\ln{\varphi} }\right)}
{\sqrt{-\ln{\Gamma/\Gqm}}}
\end{equation}
on scales $\Gamma<\Gqm$.
Asymptotically, this distribution converges to $\pG\sim 1/\Gamma$ for small $\Gamma$.
Quite importantly, the transition to this asymptotic behavior can extend over many orders of magnitude
depending on the scaling parameters $\omega$ and $\varphi$.
In order to check the perturbation prediction we constructed a quantum
graph realization (see below) of the chain model.
In fact, one finds an excellent agreement of the numerical data with Eq.~(\ref{pertgamma}),
as shown in Fig.~\ref{pogfig}a.
Regular islands, which are not included in the chain model, give an additional
contribution $\pG\sim 1/\Gamma$\cite{casa99}.
Numerically, we find for the kicked rotor the broad transition
region $(10^{-3}<\Gamma/\Delta<1)$ due to the hierarchy and the asymptotic power law
(Fig.~\ref{pogfig}a, upper inset).
%

% FCF
%
We now want to apply the concept of the flux barrier together with our findings for the break time
and the distribution of resonance widths to conductance fluctuations.
We will show that FCF come from regions before the flux barrier while isolated resonances
are scattering signatures of the hierarchical states living behind the flux barrier.
Thus, in general, they coexist but appear on different energy scales.
Since the semiclassical derivation of FCF\cite{ketz96} does not take into account the quantum
effects caused by the flux barrier, it is valid for times $t<\tqm$ only.
This predicts that FCF have a smallest energy scale
\begin{equation}\label{escale}
\Delta E^*=\Gqm=\frac{h}{\tqm}\sim\Delta\cdot\heff^{1/\gamma-1}\quad,
\end{equation}
and that they originate from phase space regions before the flux barrier.
The largest energy scale for FCF is related to the time $t_0$ at which the
classical power-law decay starts and is given by $\t0 = \Omega_{1}/\Phi_{0,1}\sim\tH\heff$
for the chain model.
The ratio $\tqm/\t0=\heff^{-1/\gamma}$ thus determines over how many orders of
magnitude FCF can be observed.
%

% isolated resonances
%
Below the energy scale $\Delta E^*$,  the conductance $g(E)$ is determined by
the states behind the flux barrier, which are characterized by $\pG$ of Eq.~(\ref{pertgamma}).
Each of the resonances will lead to an isolated feature of width $\sim\Gamma$
in the conductance.
Under the assumption that the  height and the width of these features are uncorrelated,
it was shown in Ref.~\cite{bodo00} that $\pG\sim \Gamma^{\delta-2}$ leads to $\var\sim \DE^{\delta}$.
Therefore, the asymptotic power law $\pG\sim 1/\Gamma$ should lead to a linear increase
$\var\sim\DE$ for small $\DE$ together with a broad transition region.
%

% numerics for the quantum graph
%
In order to check the above predictions for conductance fluctuations we construct
a quantum graph realization of the chain model.
Each of its regions $n$ is modelled by a fully connected graph\cite{kottos}
with $v$ vertices and a total length proportional to $\Omega_n$.
These graphs are connected such that the flux from region $n$ to $n+1$ is $\Phi_{n,n+1}$
with leads attached to region $n=0$ (Fig.~\ref{pogfig}a lower inset),
details will be published elsewhere.
By increasing $v$ the semiclassical limit, $\heff\rightarrow0$, is approached and the
flux barrier moves deeper into the hierarchy.
For  $v=32$ ($\heff=0.02$, $\omega=0.6$,$\varphi=0.21$) the flux barrier is located
between regions $n=3$ and $n=4$.
Indeed, the conductance shows FCF on energy scales above $\DE^*=14.8\cdot\Delta$ (Fig.~\ref{modelfig}b)
and isolated resonances on scales below $\DE^*$ (Fig.~\ref{modelfig}c).
When restricting this graph to regions $n\le3$, i.e., regions before the flux barrier, we find FCF
without isolated resonances.
By successively appending regions $n\ge4$ a growing number of isolated resonances appears.
For a quantitative analysis of the isolated resonances we use a quantum graph with $v=4$
$(\heff=2)$, showing many isolated resonances (Fig.~\ref{pogfig}b inset), since the flux barrier
is located between regions $n=0$ and $n=1$.
For small $\DE$ we find $\var\sim\DE$ together with a broad transition region confirming
the predictions of the perturbation theory (Fig.~\ref{pogfig}b).
%

% general discussion
%
Our analysis suggests that the numerically observed isolated resonances for the cosine billiard\cite{bodo00}
are scattering signatures of the phase space region behind the flux barrier.
The size of this region can be estimated from the relative number of isolated resonances (18\%).
From a classical simulation we determine the time up to which only regions before the flux barrier
are explored (82\% of the phase space), which gives an estimate for the break time $\tqm\approx 25$
(in units of the traversal time).
Thus the classical power law which starts at $t_0\approx 3$ (Fig.~1 in Ref.~\cite{bodo00})
can be mimicked for less than one order
of magnitude and this explains why no FCF were observed for the considered energy range.
Another puzzling observation in Ref.~\cite{bodo00} was the occurrence of the classical exponent $\gamma$
in a power law of $\var$ on scales below the mean level spacing.
Our analysis suggests, however, that in this regime one has a broad transition to the asymptotic
$\var\sim\DE$. Since this transition can extend over many orders of magnitude, it may
locally fake a power law with exponent larger than one.
Thus the numerically observed isolated resonances are explained by our theoretical approach.
%

% experiment
%
Experimentally, FCF have been observed\cite{sach98} in semiconductor nanostructures 
where typically the conductance is measured as a function of an external magnetic field.
In order to observe isolated resonances, one has to measure the conductance on magnetic field scales
smaller than the smallest scale $\Delta B^*$ of FCF.
The scale $\Delta B^* = h/(eA^*)$ originates from the area $A^*$ enclosed by a trajectory which stays
up to the break time $\tqm=h/\DE^*$ in the cavity.
If the phase coherence time $\tau_\varphi$ is larger than $\tqm$, one can observe isolated resonances.
%

% thanks
%
We thank B.~Huckestein and F.~Steinbach for helpful discussions.


\begin{thebibliography}{1}

%
\bibitem{MM74} L.~Markus and K.~R.~Meyer, {\em Generic Hamiltonian
Dynamical Systems are Neither Integrable nor Chaotic}, Memoirs of the
American Mathematical Society, No. 114 (American Mathematical Society,
Providence, RI, 1974). 
%
\bibitem{lich92} A.~J.~Lichtenberg and M.~A.~Lieberman, {\em Regular and Chaotic Dynamics},
Appl. Math. Sciences~38, 2nd ed., (Springer-Verlag, New York, 1992).
%
\bibitem{powerlaw} B.~V.~Chirikov and D.~L.~Shepelyansky,
in {\em Proceedings of the IXth Intern. Conf. on Nonlinear
Oscillations, Kiev, 1981} [Naukova Dumka {\bf 2}, 420 (1984)]
(English Translation: Princeton University Report No. PPPL-TRANS-133,
1983); C.~F.~F.~Karney, Physica {\bf 8} D, 360 (1983);
B.~V.~Chirikov and D.~L.~Shepelyansky, Physica {\bf 13} D, 395 (1984);
P.~Grassberger and H.~Kantz, Phys. Lett. {\bf 113} A, 167 (1985);
B.~V.~Chirikov and D.~L.~Shepelyansky, Phys.\ Rev.\ Lett. {\bf 82},  528 (1999).
%
\bibitem{kay84} R.~S.~MacKay, J.~D.~Meiss, and I.~C.~Percival, Physica D 13, {\bf 55} (1984);
J.~D.~Meiss, Rev. Mod. Phys. {\bf 64}, 795 (1992).
%
\bibitem{bohigas} O.~Bohigas, S.~Tomsovic, and D.~Ullmo, Phys. Rep. {\bf 223}, 45 (1993).
%
\bibitem{ketz00} R.~Ketzmerick, L.~Hufnagel, F.~Steinbach, and M.~Weiss,
Phys. Rev. Lett. {\bf 85}, 1214 (2000).
%
\bibitem{datta95} S.~Datta, {\em Electronic transport in mesoscopic systems}, 
Cambridge University Press 1995.
%
\bibitem{jala00} For a review see, e.g., R.~A.~Jalabert, Proceedings of the International School of Physics
`Enrico Fermi' Course CXLIII "New Directions in Quantum Chaos",
Edited by G. Casati, I. Guarneri and U. Smilansky, IOS Press, Amsterdam, 2000.
%
\bibitem{ketz96} R.~Ketzmerick, Phys.\ Rev.\ B {\bf 54},  10841  (1996).
%
\bibitem{hegg96} H.~Hegger {\it et~al.}, Phys.\ Rev.\ Lett. {\bf 77},  3855  (1996).
%
\bibitem{sach98} A.~S.~Sachrajda {\it et~al.}, Phys.\ Rev.\ Lett. {\bf 80},  1948  (1998);
A.~P.~Micolich {\it et~al.}, J. Phys.:Condens. Matter {\bf 10} (1998) 1339;
Y.~Ochiai {\it et~al.}, Semicond. Sci. Technol. {\bf 13} (1998) A15;
Y.~Takagaki {\it et~al.}, Phys. Rev. B {\bf 15} (2000) 10255.
%
\bibitem{ital00} G.Casati, I.~Guarneri, and G.~Maspero, Phys. Rev. Lett. {\bf 84}, 63 (2000);
%
\bibitem{taka00}Y.~Takagaki and K.~H.~Ploog, Phys. Rev. B {\bf 15} (2000) 4457;
E.~Louis and J.~A.~Verg\'es, Phys. Rev. B {\bf 15} (2000) 13014;
%
\bibitem{bodo00} B.~Huckestein, R.~Ketzmerick, and C.~Lewenkopf, Phys. Rev. Lett. {\bf 84}, 5504 (2000).
%
\bibitem{HCM85} J.~D.~Hanson, J.~R.~Cary, and J.~D.~Meiss, J. Stat. Phys. {\bf 39}, 327 (1985);
J.~D.~Meiss and E.~Ott, Phys. Rev. Lett. {\bf 55}, 2741 (1985);
T.~Geisel, A.~Zacherl, and G.~Radons, Phys. Rev. Lett. {\bf 59}, 2503 (1987).
%
\bibitem{volume} For some $\nss>\ns$ the regions $n>\nss$ will be too small to be resolved quantum
mechanically.
In the limit of small $\heff$ one has $\nss\gg\ns$ since the volumes $\Omega_n$ scale down slower
than the fluxes $\Phi_{n,n+1}$. Therefore we will concentrate on the limit $\nss\rightarrow\infty$
for simplicity.
%
\bibitem{casa99} G. Casati, G. Maspero, and D.L. Shepelyansky, Phys.\ Rev.\ Lett. {\bf 82},  524 (1999).
%
\bibitem{sund99} B.~Sundaram and G.~M.~Zaslavsky, Phys.\ Rev.\ E {\bf 59}, 7231 (1999).
%
\bibitem{kottos} T.~Kottos and U.~Smilansky, Phys. Rev. Lett. {\bf 79}, 4794 (1997);
Phys. Rev. Lett. {\bf 85}, 968 (2000).

\end{thebibliography}
\end{document}